\documentclass[twocolumn,aps,prb]{revtex4}  

\usepackage{epsfig}

\begin{document}

\title{Observation of disorder-induced weakening of electron-phonon interaction in thin noble metal films}

\author{ J. T. Karvonen} 
\author{ L. J. Taskinen}
\author{I. J. Maasilta}

\affiliation{Nanoscience Center, Department of Physics, 
P.O. Box 35, FIN-40014 University of Jyv\"{a}skyl\"{a}, Finland}

\date{\today}

\begin{abstract}

We have used symmetric normal metal-insulator-superconductor 
(NIS) tunnel junction pairs, known as SINIS structures, for 
ultrasensitive thermometry in the temperature range 50 - 700 
mK.  By Joule heating the electron gas and measuring the 
electron temperature, we show that the electron-phonon (e-p) scattering 
rate in the simplest noble metal disordered thin films 
(Cu,Au) follows a $T^4$ temperature dependence, leading to a stronger decoupling of the electron gas from the lattice 
at the lowest temperatures. This power law is indicative e-p coupling mediated by vibrating disorder, in contrast to 
the previously observed $T^3$ and $T^2$ laws.

\end{abstract}


\maketitle

Although the interaction between conduction electrons and thermal phonons is elementary for many processes and 
phenomena at low temperatures, there are still relatively few experimental studies that conclusively support the theoretical 
description, particularly for typical disordered thin film samples.
Several earlier results \cite{rou,kansk,urbina} indicated that even for disordered films, the temperature dependence 
for the electron-phonon (e-p) scattering rate $1/\tau_{e-p}$ follows the power law expected for pure samples with 
coupling to longitudinal phonons only \cite{gant}: $1/\tau_{e-p} \sim T^3$. These results confirmed the relation 
$P=\Sigma \Omega(T_e^5-T_p^5)$ between heating power $P$ and electron and phonon temperatures $T_e$ and $T_p$ in a volume $\Omega$, and it is widely 
used for thin film metallic samples at low temperatures ($\Sigma$ is a material dependent parameter).
However, the theory for  disordered thin films \cite{schmid,alt,reiz,serg} predicts that the scattering rate from vibrating disorder (impurities, 
boundaries etc.) is $1/\tau_{e-p} \sim T^4$ in the limit $ql < 1$, where $q$ is the wavevector of the dominant thermal phonons and  
$l$ the electron mean free path. This leads to the relation 

\begin{equation} 
P = \Sigma'\Omega (T_e^6-T_p^6),
\label{eq}
\end{equation}

a result that has not been widely confirmed. In fact, we are not aware of any observation of it in standard normal 
metal films like Cu, Al, Au etc., and some evidence exists only for strongly disordered ($l \sim 1$ nm) Ti, Hf \cite{ger}
and Bi \cite{kom} films.  

Here, we report the observation of disorder-weakened electron-phonon (e-p) scattering in ordinary evaporated
Cu and Au thin films. We have measured the rate at which electrons in a normal metal wire overheat, when DC power
is applied to it by Joule heating, at sub-Kelvin temperatures \cite{jenni}. This technique has been shown 
\cite{rou,kansk,urbina} to give the energy-loss rate directly,
in contrast to the temperature dependence of the weak localization resistance \cite{lin}, which gives the dephasing 
rate \cite{depha}. The overheating rate is determined by measuring the electron temperature $T_e$ directly with the help of symmetric 
normal metal-insulator-superconductor (NIS) tunnel junction pairs, known as SINIS structures. SINIS-thermometers
have been shown \cite{row,nah1,leivo} to be extremely sensitive thermometers, operating at the lowest experimentally 
achievable temperatures. Therefore, they are also candidates for ultrasensitive microbolometers \cite{Nahum,Kuzmin} in the
sub-Kelvin temperature range.

All the samples used in this work have two normal metal wires of length  $\sim$ 500 $\mu$m,
and width $\sim$ 300-400 nm, separated by a distance 2 $\mu$m and electrically isolated from each other, as shown in 
the SEM image [Fig. 1(a)]. They were fabricated by e-beam lithography and three-angle shadow-mask evaporation
technique on thermally oxidized (oxide thickness 250 nm) Si substrates, with film thicknesses varying from 33 nm to 
140 nm, corresponding to a measured $l=$ 24-53 nm. The deposition was done by e-beam evaporation in a UHV chamber with
a growth rate $\sim 1-2$ \AA/s.

The normal metal wires were connected to the measurement circuit by superconducting Al leads. 
One pair of leads forms SINIS tunnel junctions for each wire, and are used to measure
the electron temperature of the normal metal wire. These tunnel junctions were formed by thermal oxidation of the Al 
films, yielding tunneling resistances $R_T \sim$ 20-50 k$\Omega$, typically. 
In contrast, the junctions connecting the lower wire in Fig. 1 to a voltage source are NS-junctions without tunneling 
barriers, and are used to heat the sample. Clean NS-junctions have a small electrical resistance compared to the
residual low-temperature resistance of the wire, between 100 and 1500 $\Omega$ for all the samples.
 For all heating voltages used in this paper, we have measured that the NS-junctions are biased within the 
superconducting
gap $\Delta$ of the leads. Then, the process of Andreev reflection can take place, making the junctions good
thermal insulators despite being electrically conducting. This way we can achieve conditions in which the wire is 
uniformly Joule heated \cite{short}. Moreover, since the wires are much longer than a typical electron-electron 
scattering length \cite{hug}, the electrons in the wire have a well defined Fermi-distribution, and a temperature 
$T_e$. In addition, the effect of multiple Andreev reflections on the distribution is negligible for these wire 
lengths \cite{anne}.
Also, the thermal resistance of the tunnel junctions is 
estimated to be at least an order of magnitude larger than the thermal resistance due to the e-p coupling $R_{e-p}$ at the 
temperature range of this 
experiment\cite{dragos}, and can be neglected in the analysis.

All measurements were performed by current biasing the two SINIS thermometers and measuring their DC voltages 
simultaneously [Fig 1(b)] 
in a dilution refrigerator with a base temperature $\approx 60$ mK. All wiring connected to the sample was filtered 
at 4.2 K, to minimize the overheating by external noise. This was seen to be very critical for good measurements at 
the lowest temperatures, as we also performed measurements without the filtering, and observed significant overheating 
below 0.3 K. 
The SINIS thermometers were calibrated by varying the bath temperature ($T_{bath}$) of the refrigerator 
very slowly, to ensure the sample stage was in equilibrium with the bath. 
The temperature of the sample stage was measured with a calibrated Ruthenium Oxide thermometer.

Figure 2 shows an example of a calibration measurement.
The theoretical points (red) were calculated numerically from the BCS theory.
This calculation has essentially no free parameters, since the current bias $I$ is known, and $R_T$ and $\Delta$ are 
determined independently
from the I-V characteristics (not shown). The agreement with the data is good 
except at the very lowest temperatures, where a slight deviation can be seen below $T_{bath}=$0.1 K. This deviation 
could arise by at least two major mechanisms: (i) a component of leakage or noise current
affects the measurement, and (ii) the electrons overheat at the low-$T$ range ($T_e > T_{bath}$). We have ruled out 
(i) by directly measuring the 
responsivities $dV/dP$ of the thermometers as a function of $I$ using a lock-in amplifier (Fig. 2 inset). The 
theoretical $dV/dP$ is a monotonically decreasing function of $I$. However, experimentally at low $I$ the responsivity deviates from theory, and we 
find a maximum. Thus, by setting the bias current above the maximum ($I \ge 60$ pA in Fig. 2 inset), where the BCS-theory is accurate, 
problem (i) can be avoided. 
Mechanism (ii), overheating by noise power $P_{noise}$ is an obvious explanation, since we observed a clear 
improvement with added filtering of the lines. Thus, at the lowest temperatures ($< 0.1$ K) the SINIS thermometers 
start to saturate to a temperature $T_e$ determined by the absorbed noise power $P_{noise}\propto T_e^n-T_p^n$, where 
$n=5,6$ depending on the mechanism of e-p scattering involved. The correct $T_e(V)$ is therefore always given by the 
BCS-theory function, instead of the measured phenomenological $V(T_{bath})$ plot.    

 Using the calibrated SINIS thermometers, we then studied what happened when only the lower wire is directly
heated by Ohmic dissipation at $T_{bath}\approx$ 60 mK. Heating power $P$
was applied by slowly ramping a DC voltage across the wire, and determined
by measuring the heating current and voltage directly in a 4-wire configuration. In all the samples studied, the 
temperature of the upper wire rises also, when the lower wire is heated \cite{jenni}. This indirect heating can be 
either due to phonons emitted from the hot wire and backscattered, or by photons (thermal noise) coupling 
capacitively. Since we don't know the relative contributions of these two channels, we only obtain an upper limit for 
the local phonon temperature \cite{fooot} $T_p$. However, the main conclusion is that all the data satisfy the 
condition $T_e^n \gg T_p^n$ for powers $n=5,6$, and the error introduced by approximating Eq. (\ref{eq}) by $P = 
\Sigma'\Omega T_e^6$ is less than 4\% (for $n=5$ 6\%). Thus, analysis can proceed without the knowledge of $T_p$, and 
the slope of $T_e$ vs. $P$ in a log-log plot will directly give $1/n$. We also note that thermal relaxation by 
radiation into the leads \cite{sch} is estimated to be insignificant compared to the e-p scattering, due to the large 
volume of our samples.

In Fig. 3 we plot the electron temperatures in the heated wire vs. $P$ obtained from several sweeps for four different 
samples in log-log scale. Data for the thinner Cu and Au samples is shown in Fig. 3(a), whereas Fig. 3(b) plots the 
results for the thicker Cu samples. The agreement with $n=6$ is strikingly good for the thin Cu and Au samples, from 
$\sim 0.1$ pW to $\sim 1$ nW. The 84 nm thick Cu sample also agrees with $n=6$ at higher power, but the thickest Cu 
sample does not, and is consistent with $n=5$. The saturation of $T_e$ by $P_{noise}$ at the lowest $P$ is clearly 
visible for all the samples, and the effect seems to be stronger for the thicker Cu samples. This is most likely due 
to the changing resistance of the wire: the thicker wires have lower resistance approaching the characteristic 
impedance of the wiring, causing more absorption of $P_{noise}$.

To obtain a more detailed picture, we also plot the numerically computed logarithmic derivatives of the data vs. $P$ 
in Fig. 4. Even though numerical differentiation is sensitive to noise, we can still strengthen our case. Again, for 
the thinner samples $n=6$ is obviously more consistent than any other integer power. Some deviations start to appear 
above a few hundred pW, which are not fully understood at the moment. 
The thicker samples show the effect of $P_{noise}$ by a smooth increase of the derivative at low powers. However, the 
84 nm thick sample does seem to develop a plateau at around $n=6$, whereas the thickest sample is more consistent with 
$n=5$ in the high power limit.

By determining the volumes of the normal metal wires accurately with AFM and SEM measurements, we can extract the 
parameters $\Sigma'$ and $\Sigma$ from the one parameter fits in Fig. 3. We get for the Cu samples $\Sigma'= 2 \times 
10^{10}$ W/K$^6$m$^3$ ($t=33$ nm), $\Sigma'= 6 \times 10^{9}$ W/K$^6$m$^3$ ($t=84$ nm) and $\Sigma= 3 \times 10^{9}$ 
W/K$^5$m$^3$ ($t=140$ nm), whereas for the Au sample $\Sigma'= 6 \times 10^{10}$ W/K$^6$m$^3$. We note that the value 
of $\Sigma$ in the thickest Cu sample is consistent with earlier studies \cite{Nahum}, since it is a material 
parameter in the pure limit. However, theory \cite{serg} predicts that $\Sigma'$ depends on the level of disorder, and 
thus if $ql < 1$, each sample can have a different $\Sigma'$. This is clearly the case in our experiment. Comparing 
the measured $\Sigma'$ with the microscopic theory, we can estimate the transverse sound velocity $c_t$. We get $c_t= 
1800$ m/s and $c_t=2700$ m/s for the Cu wires $t=33$ nm and $t=84$ nm, respectively, and $c_t= 1200$ m/s for the Au 
wire, a reasonable result. As the bulk longitudinal sound velocities in Cu and Au are about twice as large, we have 
indirect evidence that transverse modes are dominant for $ql<1$.

In conclusion, we have for the first time obtained clear evidence that the electron-phonon scattering rate scales with 
temperature as $1/\tau_{e-p} \sim T^4$ in  disordered, evaporated Cu and Au thin films. This power law corresponds to 
electrons scattering from phonons mediated by vibrating disorder. In our measurement boundary scattering is likely to 
be significant, since $l\sim t$. In contrast, e-p scattering in the presence of static disorder leads to $1/\tau_{e-p} 
\sim T^2$, a result that has been confirmed in many materials and samples \cite{lin}. Our result has several practical 
consequences. Compared to a lower power dependence, it is harder to cool the electrons with cold phonons, i.e. the 
electron gas decouples from the lattice more strongly at the lowest temperatures. Direct electron cooling 
\cite{nah1,leivo} then becomes more important. On the other hand, a sensor based on the hot electron effect is even 
more sensitive, since the ultimate noise equivalent power of such a detector is proportional to $1/R_{e-p}^{1/2}$. We 
stress that in order to obtain $1/\tau_{e-p} \sim T^4$ the sample needs to be a regular metal, have $ql < 1$, contain 
little non-vibrating disorder and have 3D phonons coupled to electrons by the deformation potential. We believe most 
earlier experiments do not satisfy all these conditions (e.g. Ref. \cite{rou} has $ql > 1$, Ref. \cite{yung} has a 
suspended GaAs substrate, Refs. \cite{kansk,urbina,ditusa} studied alloys and Ref. \cite{kivin} studied degenerately 
doped Si in the strong screening limit.)

     We thank D.-V. Anghel, A. N. Cleland, J. P. Pekola, H. Pothier, A. Savin and A. Sergeev for discussions.  This 
work was supported by the Academy of Finland under the Finnish Center of Excellence 
Program 2000-2005 project No. 44875, and by the projects No. 105258 and 205476 (TULE programme).

\begin{figure*}[h]

\centerline{\epsfig{file=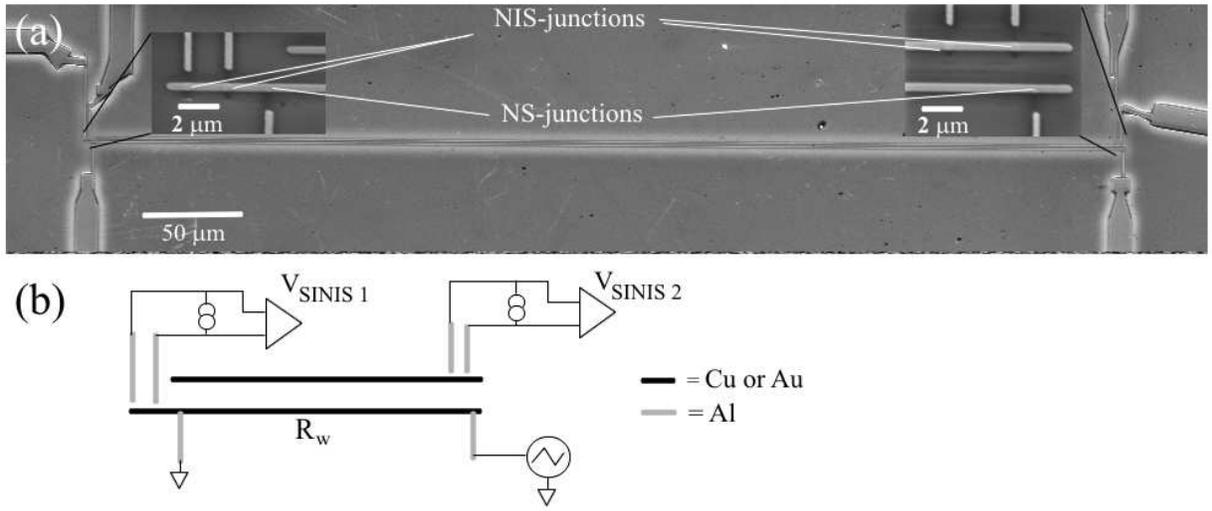,width=0.9\linewidth,angle=0}}

\caption{(a) A SEM image of a sample. The horizontal lines are Cu wires, the vertical lines are Al 
leads that form the junctions at the intersections with the Cu line. The insets are enlargements of the 
areas where the junctions are located. Note that the pure Al vertical lines (black) make contact
with the horizontal Cu lines, the lighter gray vertical lines consist of Al+shadows from the Cu 
evaporation.  (b) A schematic of the sample and the measurement circuit. SINIS
bias circuit is fully floating with respect to the heating circuit, thus no heating current flows through the NIS 
junctions at DC.}

\end{figure*}

\begin{figure}[h]

\centerline{\epsfig{file=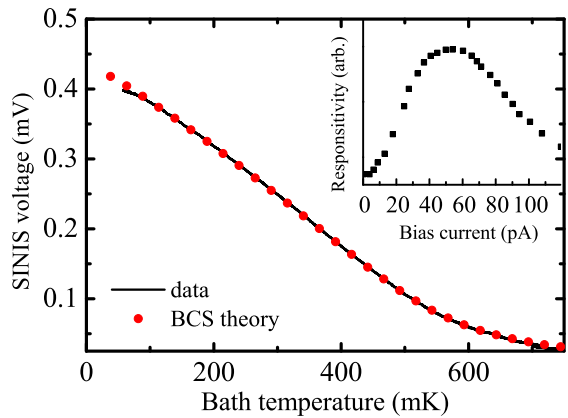,width=\linewidth,angle=0}}

\caption{(color online) Calibration data for a typical SINIS thermometer biased at $I=60$ pA.
Black line is the experimental data, 
and red dots the BCS-theory result.  
For the BCS-theory points, $\Delta$ and $R_T$ were determined independently from the experimental I-V characteristics. 
From this data we estimate the temperature sensitivity $\delta T=(dT/dV) \delta V \approx $0.1 mK rms, where $\delta 
V$ is the rms noise voltage. Inset: Measured responsivity 
$dV/dP$ of a SINIS vs. the bias current.} 

\end{figure}

\begin{figure}[h]

\centerline{\epsfig{file=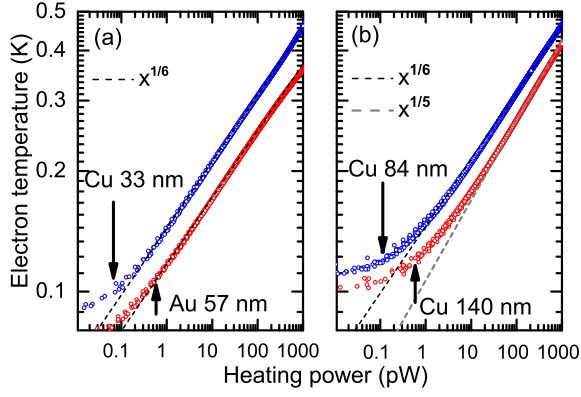,width=\linewidth,angle=0}}

\caption{(color online) The electron temperature of (a) $t=33$ nm Cu wire (blue) and $t=57$ nm Au wire (red) vs. 
applied Joule heating power in log-log scale. (b) Same for $t=84$ nm (blue) and $t=140$ nm (red) Cu wires. Dashed 
black lines are fits of the form $T=(P/A)^{1/6}$, corresponding to the disordered e-p scattering theory [Eq. (\ref{eq})]. Dashed gray line is a fit 
$T=(P/A)^{1/5}$. $T_{bath} =$60 mK.}

\end{figure}

\begin{figure}[h]

\centerline{\epsfig{file=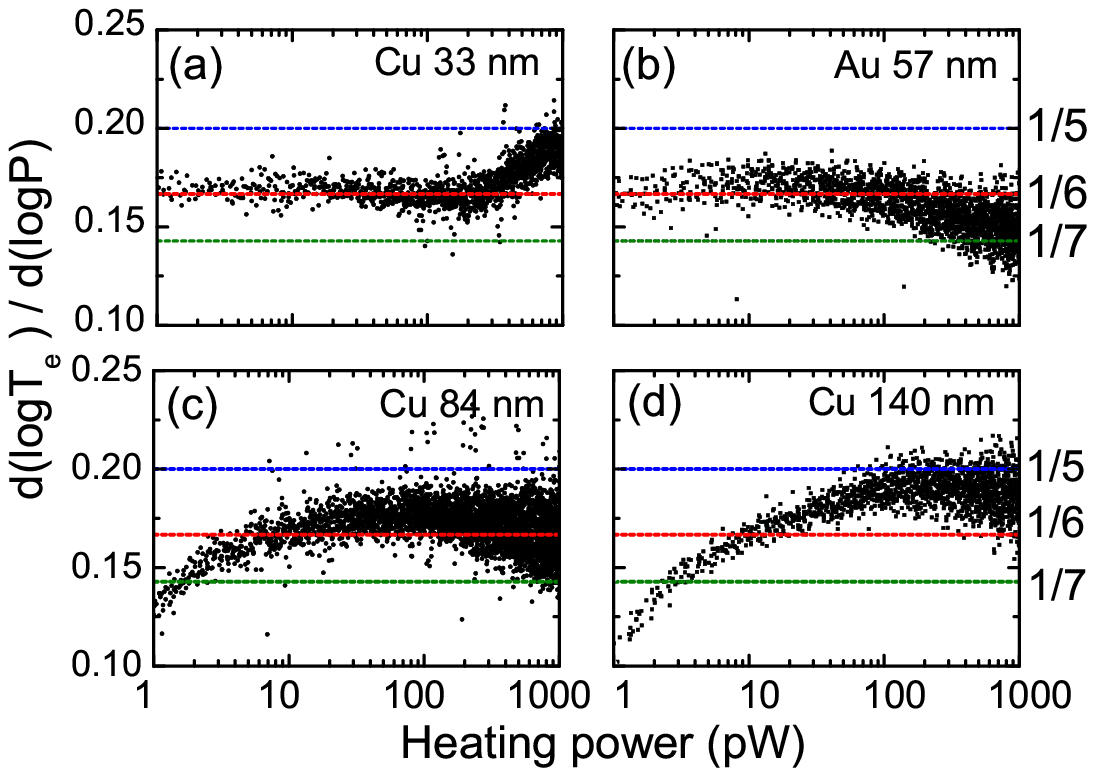,width=\linewidth,angle=0}}

\caption{(color online) The logarithmic numerical derivative $d(\log T_e)/d(\log P)$ of the same data as in Fig. 3 
(points). (a) Cu $t=33$ nm, (b) Au $t=57$ nm, (c) Cu $t=84$ nm, and (d) Cu $t=140$ nm. Blue line equals 1/5, red 1/6 
and green 1/7. }

\end{figure}


\begin{thebibliography}{}
\bibitem{depha}  The dephasing rate can equal the energy loss rate if the electron-phonon interaction is the only 
dephasing mechanism present. In general, one can 
imagine many other dephasing channels that do not influence the energy loss rate, a prime example being the 
electron-electron interaction, which by definition cannot lower the energy of the electron gas as a whole.
\bibitem{short} For much shorter but otherwise equivalent wires, the NS junctions are quickly biased above the gap, 
and the heat leak through them has to be taken into account in the analysis \cite{kivi}.
\bibitem{fooot} Since the two wires are separated by only $\sim 2 \mu$m, a distance much shorter than the lateral mean 
free path of phonons in SiO \cite{swar}, it is reasonable to assume that the phonon temperatures are equal in the two 
wires.
\end{thebibliography}
\end{document}